**JSWSC**

RESEARCH ARTICLE      OPEN ∂ ACCESS

# A homogeneous *aa* index: 2. Hemispheric asymmetries and the equinoctial variation

Mike Lockwood[1,*], Ivan D. Finch[2], Aude Chambodut[3], Luke A. Barnard[1], Mathew J. Owens[1], and Ellen Clarke[4]

[1] Department of Meteorology, University of Reading, Whiteknights Campus, Earley Gate, PO Box 243, Reading RG6 6BB, UK
[2] Rutherford Appleton Laboratory, Chilton, Didcot, OX11 0QX, Oxfordshire, UK
[3] Institut de Physique du Globe de Strasbourg, UMR7516, Université de Strasbourg/EOST, CNRS, 5 rue René Descartes, 67084 Strasbourg Cedex, France
[4] British Geological Survey, Edinburgh EH14 4AP, UK

**Abstract** – Paper 1 (Lockwood et al., 2018) generated annual means of a new version of the *aa* geomagnetic activity index which includes corrections for secular drift in the geographic coordinates of the auroral oval, thereby resolving the difference between the centennial-scale change in the northern and southern hemisphere indices, $aa_N$ and $aa_S$. However, other hemispheric asymmetries in the *aa* index remain: in particular, the distributions of 3-hourly $aa_N$ and $aa_S$ values are different and the correlation between them is not high on this timescale ($r = 0.66$). In the present paper, a location-dependant station sensitivity model is developed using the *am* index (derived from a much more extensive network of stations in both hemispheres) and used to reduce the difference between the hemispheric *aa* indices and improve their correlation (to $r = 0.79$) by generating corrected 3-hourly hemispheric indices, $aa_{HN}$ and $aa_{HS}$, which also include the secular drift corrections detailed in Paper 1. These are combined into a new, "homogeneous" *aa* index, $aa_H$. It is shown that $aa_H$, unlike *aa*, reveals the "equinoctial"-like time-of-day/time-of-year pattern that is found for the *am* index.

**Keywords:** Space climate / Space weather / Geomagnetism / Space environment / Historical records

## 1 Introduction

### 1.1 The *aa* and *am* indices

As discussed in Paper 1 (Lockwood et al., 2018), the *aa* index was devised by Mayaud (1971, 1972, 1980) to give a continuous, well-calibrated and homogeneous record of geomagnetic activity that extends back to 1868. It uses just two stations at similar geomagnetic latitudes, one in each hemisphere, and averaging the data from them, to a large extent, gives cancellation of the seasonal variation in the geomagnetic response to solar forcing that is seen at either one of the stations individually. Figure 1a shows that this is effectively achieved for the "classic *aa*" (i.e., the official *aa* index generated by EOST (École et Observatoire des Sciences de la Terre), as available from ISGI (International Service of Geomagnetic Indices, http://isgi.unistra.fr/) and other data centers around the world. The *aa* indices show the well-known semi-annual variation in geomagnetic activity (Cortie, 1912; Chapman & Bartels,

1940; Cliver et al., 2002; Le Mouël et al., 2004), with equinoctial peaks in average values: this can be seen in Figure 1a for the northern hemisphere index, $aa_N$ (in red), for the southern hemisphere index, $aa_S$ (in blue) and for the average of the two, *aa* (in black). The average annual variation is, however, different in $aa_S$ and $aa_N$, such that in northern-hemisphere winter (i.e., around time-of-year $F = 0$ which is defined to be at midnight between 31 December and 1 January and so is also $F = 1$), $\langle aa_N \rangle$ and $\langle aa_S \rangle$ (and therefore $\langle aa \rangle$) are very similar. However, in northern-hemisphere summer ($F$ around 0.5), $\langle aa_N \rangle$ is considerably greater than $\langle aa_S \rangle$. This difference is averaged out in $\langle aa \rangle$, such that only in $\langle aa \rangle$ is the December minimum the same depth as the June minimum, showing that the annual and seasonal variations have been averaged out leaving only the semi-annual variation, with its peaks near the equinoxes.

All plots in Figure 1 are restricted to data from the years 1959–2017 so that they can be compared to the *am* indices, the average variations of which are shown in the second row of the figure. The *am* index (Mayaud, 1980) is, like *aa*, a 3-hourly range index but compiled using area-weighted means of data from rings of mid-latitude stations, currently with 11 in the northern hemisphere and 10 in the southern. It is compiled by ISGI (and collaborating institutes) who make available the northern hemisphere index, *an*, the southern hemisphere index, *as*, and *am* = (*an* + *as*)/2 for 1959 to the present day. The annual variations of ⟨*am*⟩, ⟨*an*⟩ and ⟨*as*⟩ are shown in Figure 1c (in black, red and blue, respectively), which shows that the behaviour is very similar indeed to that for the *aa* indices in Figure 1a. Therefore, at least in terms of its variation over the year, the *aa* index certainly succeeds in its aim of replicating an equivalent index derived using a more extensive array of observatories.

*Corresponding author: m.lockwood@reading.ac.uk

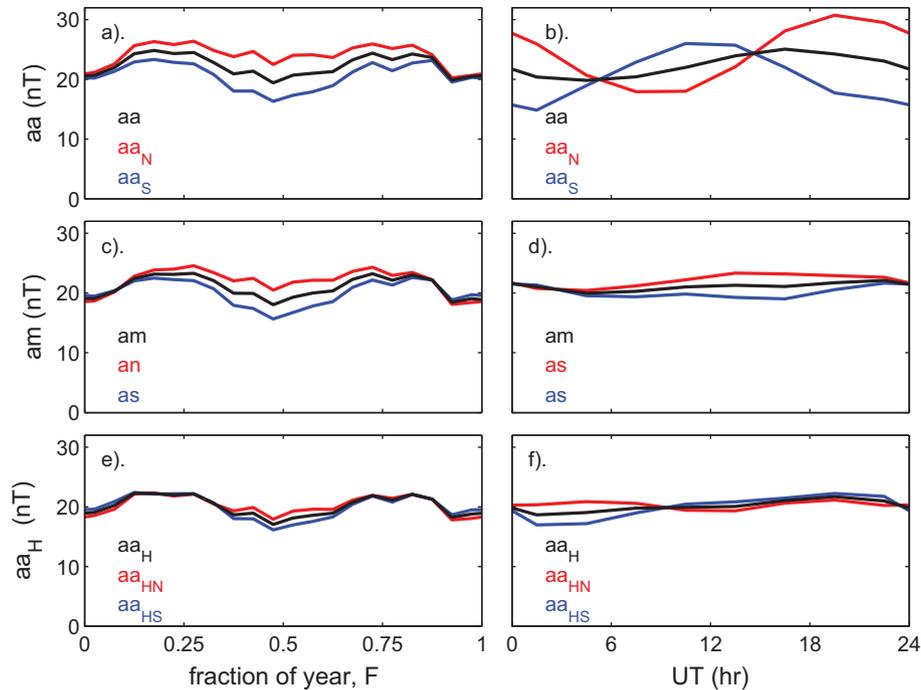

**Fig. 1.** Annual (left column) and diurnal (right column) variations in mean geomagnetic indices for 1959–2017. In all panels, the red line shows the northern hemisphere index, the blue line is the corresponding southern hemisphere index and the black line is the global index, being the average of the two hemispheric indices. (a) and (b) are for the classic *aa* indices with $aa_N$ in red, $aa_S$ in blue and *aa* in black. (c) and (d) are for the *am* indices with *an* in red, *as* in blue and *am* in black. (e) and (f) are for the new homogenized *aa* indices derived in this paper with $aa_{HN}$ in red, $aa_{HS}$ in blue and $aa_H$ in black.

The two *aa* stations are also roughly 10 h apart in local time and it was hoped in the construction of *aa* that this would largely cancel out the diurnal variation at the two stations. Comparison of Figures 1b and 1d shows that this is considerably less well achieved for *aa* than it is for *am*. The distribution of *am* stations with longitude in each hemisphere is not ideal which introduces a small spurious UT variation; however, this is very much smaller than for *aa* which has only one station in each hemisphere. The diurnal variations in ⟨*aa*_N⟩ and ⟨*aa*_S⟩ are not quite in antiphase, nor are they exactly the same in amplitude or waveform: as result, ⟨*aa*⟩ shows considerable average diurnal variation (Fig. 1b). On the other hand, the use of rings of longitudinally-spaced stations to construct *am* has suppressed the diurnal variations in both ⟨*an*⟩ and ⟨*as*⟩ (Fig. 1d) such that average *am* is almost constant with UT.

## 1.2 Time-of-day/time-of-year response patterns

The mean values for a given time-of-year (*F*) in the left-hand plots of Figure 1 are averaged over all times of day

(UT), and the mean values at a given UT in the right-hand plots are averaged over all *F*. In general, we are concerned with the full time-of-day/time-of-year (UT-*F*) patterns of variation of the geomagnetic responses. The top row of Figure 2 shows the three main UT-*F* patterns predicted from geometric considerations of solar-terrestrial interactions and the bottom row of Figure 2 shows an example of each type of pattern, as seen in averages of observations, either in near-Earth space or in geomagnetic activity (after Lockwood et al., 2016).

All three patterns arise from the geometrical considerations associated Earth's orbit around the Sun, combined (in the first two cases, at least) with the effects of Earth's rotation. The "Russell-McPherron" (R-M) pattern (Fig. 2a) arises from considering the changes in the angle between the GSM (Geocentric Solar Magnetospheric) and GSE (Geocentric Solar Ecliptic) reference frames (Russell & McPherron, 1973); the equinoctial pattern (Fig. 2b) arises from considering the angle between the solar wind direction and Earth's magnetic axis (Bartels, 1925; McIntosh, 1959) and the axial pattern (Fig. 2c) arises from the variation in Earth's heliographic latitude (Cortie, 1912) and also from the annual variation of the angle between the heliocentric Radial-Tangential-Normal (RTN) and geocentric GSE reference frames (Lockwood et al., 2016). All three predict peaks in geomagnetic activity at or near the equinoxes (but different UT dependencies). Figure 2d demonstrates that the R-M effect is seen in the average (half-wave rectified) southward component of the IMF in the GSM frame (O'Brien & McPherron, 2002), which is well understood to be the main driver of geomagnetic indices. However, neither of the geomagnetic indices shown in Figure 2, *am* and *Dst*, display the R-M pattern. The idea behind axial effect is that near the equinoxes, Earth is at slightly higher heliographic latitudes, which increases the probability of it leaving the streamer belt and encountering the fast solar wind (Hundhausen et al., 1971), especially at solar minimum (McComas et al., 2008): hence in this case there is no effect of Earth's rotation and so no UT variation. There is a second annual geometric effect associated with the variable difference between the GSE and heliocentric RTN reference frames: this effect is in antiphase with the heliographic latitude effect, favouring solstices over the equinoxes in terms of giving southward IMF and hence geomagnetic activity. It also has no UT variation but is relatively small. The axial effect appears to be present in the *Dst* index (as shown in Fig. 2f, where *Dst* has been corrected for the longitudinal inhomogeneity in the ring of equatorial stations using the procedure of Takalo & Mursula, 2001). However, Lockwood et al. (2016) point out that *Dst* is not responding to the variation in Earth's heliographic latitude, rather the long duration of large *Dst* responses (storms) to southward IMF (in the GSM frame) smooths out the UT variations seen in Figure 2d, giving an axial-like behaviour.

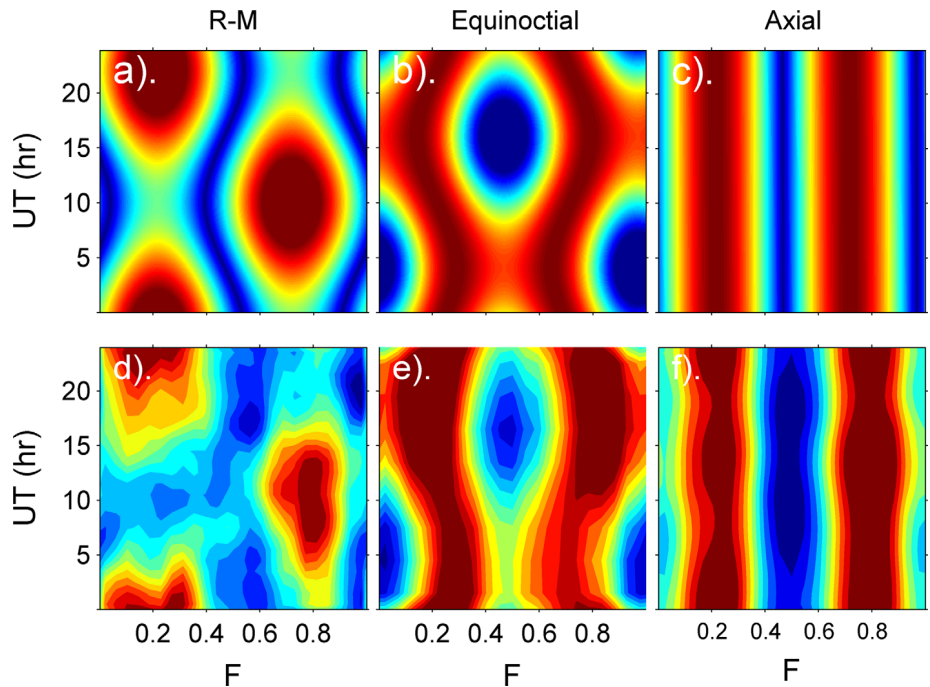

**Fig. 2.** Predicted and observed time-of-day/time-of-year patterns. The upper panel shows predictions based on geometric factors and the lower panel the contoured patterns in means of observations: (a) the Russell-McPherron (R-M) pattern; (b) the equinoctial pattern; and (c) the axial pattern (see text for details). In all plots red areas are maxima and blue are minima. (d) The R-M pattern is seen in the mean of the observed southward IMF in GSM coordinates, $B_S(GSM)$ (where $B_S(GSM) = -B_Z(GSM)$ when $B_Z(GSM) < 0$ and $B_S(GSM) = 0$ when $B_Z(GSM) \geq 0$). (e) The equinoctial pattern is seen in the *am* geomagnetic index. (f) The axial pattern is seen in the corrected *Dst* index, $-k(UT) \times Dst$, where the factor $k(UT)$ is the modelled response of the 4 *Dst* stations that allows for the fact that they are not equally spaced in longitude. In all cases we are concerned only with the form of the pattern rather than the amplitude and colour scales have been auto-scaled between the maximum and minimum values.

The UT-*F* pattern seen in the *am* index in Figure 2e has similarities to the equinoctial pattern in Figure 2b, although it is not an exact match and there are elements of all three patterns in the *am* response (Cliver et al., 2000; Chambodut et al., 2013). The equinoctial element indicates that the tilt of the Earth's rotational and/or magnetic axes towards or away from the Sun has an influence, introducing differences between the two solstices and between 4 UT and 16 UT which are not predicted by the R-M effect (O'Brien & McPherron, 2002). There have been a number of explanations proposed for this observed equinoctial pattern. These include tilt-induced changes in the ionospheric conductivity within the nightside auroral electrojet of substorm current wedge and the postulate (as yet unproven and somewhat counter-intuitive) that electrojet currents are stronger when conductivities caused by solar extreme ultraviolet (EUV) are low in both midnight-sector auroral ovals (Lyatsky et al., 2001); tilt influence on the magnetopause reconnection voltage (Crooker & Siscoe, 1986; Russell et al., 2003); the effect of tilt on the proximity of the ring current and auroral electrojet (Alexeev et al., 1996); and tilt effects on the stability of the cross-tail current sheet (Kivelson & Hughes, 1990; Danilov et al., 2013). Finch et al. (2008) used a global network of geomagnetic stations to show that the equinoctial behaviour originates during substorm expansion phases and in the substorm current wedge and is not a feature of dayside currents and flows during the substorm growth phase. (These authors showed that the dayside currents do not depend on UT and vary only with season, being greater in summer when conductivities are higher). The results of Finch et al. (2008) therefore strongly support the explanations of the equinoctial effect invoking nightside magnetospheric or ionospheric effects rather than those that postulate modulation of the magnetopause reconnection voltage. Note also that indices influenced by the substorm current wedge also depend on the solar wind dynamic pressure $P_{SW}$ ($= m_{SW}N_{SW}V_{SW}^2$, where $m_{SW}$ is the mean ion mass, $N_{SW}$ the number density and $V_{SW}$ the speed of the solar wind), because it compresses the near-Earth geomagnetic tail and so modulates the near-Earth cross-tail current there for a given open magnetic flux content in the tail (Lockwood, 2013): Finch et al. (2008) showed that a $V_{SW}^2$ dependence was present in the equinoctial pattern response but not in the directly-driven dayside response. As discussed in Paper 1, mid-latitude range indices respond primarily to the substorm current wedge, and so the results of Finch et al. (2008) explain why it is the $am$ index that displays the equinoctial pattern most clearly.

### 1.3 The aims of the present paper

In the present paper, we employ the concept introduced by Finch (2008) of the sensitivity $S_o$ of a mid-latitude geomagnetic observatory to solar wind forcing, which depends on its location (geomagnetic and geographic latitudes), its Magnetic Local Time (MLT) (and hence the UT) and on the time of year, $F$. Our motivation is to remove effects caused by the geomagnetic and geographic coordinates of the site and so homogenise the $aa$ index on sub-annual timescales, such that $aa_N$ and $aa_S$ are more highly correlated and have distributions of values that are more alike.

This also allows us to evaluate how the equinoctial time-of-year time-of-day pattern should appear in the $aa$ data once the station location effects are accounted for. Chambodut et al. (2013) have mapped the $am$ index data into 4 MLT sectors and they show that the equinoctial pattern is present in the $am$ data from each one. It is weakest in the noon sector, particularly at 0–9 UT. However, that it can be detected at such a wide range of MLT and UTs, indicates that some information about the equinoctial variation should be available in the 2-station $aa$ index, if the spurious diurnal variation caused by having only one station in each hemisphere can be removed. We develop Finch's numerical model of the stations' sensitivities by comparing the time-of-day/time-of-year pattern of response for various stations to that of the $am$ index. Our aim (as in Paper 1) is to reduce the difference between the hemispheric $aa$ indices on 3-hourly and daily timescales so that we have greater confidence that the average of the two is a representative index of the response of the global magnetosphere-ionosphere-thermosphere system to events of enhanced solar wind forcing. This would make the 150-year record of major storms from $aa$ data much more reliable and give a more reliable rank order of the severity of major geomagnetic disturbance events. As a test of this, in the present paper we study the extent to which $aa$ can reproduce the equinoctial variation that is found in equivalent range indices from more extensive and evenly-distributed networks of observatories. We show that allowing for this modelled station sensitivity can (along with the long-term recalibration described in Paper 1) effectively remove the spurious diurnal variation and known hemispheric asymmetries in the classic $aa$ index and reveals the equinoctial pattern in the $aa$ index. The bottom row in Figure 1 shows the corresponding averages of the new, "homogenized" $aa$ indices ($aa_{HN}$, $aa_{HS}$, and $aa_H = (aa_{HN} + aa_{HS})/2$) that are developed in the subsequent sections of the present paper. It can be seen that the difference in the average annual variation of $aa_N$ and $aa_S$ has almost been eliminated (leaving only a small seasonal variation with summer means slightly greater than winter ones at both solstices and not just around the June solstice), as has most of the difference in their average diurnal variations, such that average $aa_H$ is almost independent of UT, even though it is compiled from just two stations.

## 2 Methodology

Finch (2008) introduced the concept of the location-dependent magnetometer station sensitivity, $S_o$, defined for a given type of single-station geomagnetic activity measure by

$$S_o = G_A / I_S, \qquad (1)$$

where $G_A$ is the geomagnetic activity measure in question and $I_S$ is a measure of the input solar forcing, which includes the effects of both induced currents in near-Earth space and of conductivity changes due to variations in the ionizing EUV and X-ray radiations from the Sun or particle precipitations. Finch considered $S_o$ to be a function of the instrument co-ordinates only because instrument and local site characteristics are accounted for by other inter-calibration procedures. By taking ratios of $G_A$ seen simultaneously at many pairs of different stations, the $I_S$ factor is cancelled and the ratios of the station sensitivities are known. If the data from different stations are combined into a geomagnetic index using linear mathematics (such as taking an average) then the sensitivities are similarly combined. From comparisons of these ratios for many pairs of stations, Finch (2008) derived a functional form for computing the sensitivity of a station as a function of its geographic coordinates, date, time-of-year and time-of-day:

$$S_o = \{1 + A\cos^{0.7}(\chi)\}\{B\cos[(T - T^*)(2\pi/24)] + 1\}/m, \qquad (2)$$

where

$$T^* = 1.5\sin\{2\pi(F + F_1)\} - 0.5, \qquad (3)$$

$A$ and $B$ are constants, $\chi$ is the solar zenith angle, $T$ is the MLT of the station (in hours), $F$ is the fraction of the year and $F = F_1$ at the spring equinox (taken to be 100/365.25 for the northern hemisphere and 283/365.25 for the southern hemisphere). Lastly, $m$ is a normalising factor that ensures that the average value of $S_o$, over all times-of-day (UT) and all times-of-year ($F$), is unity for a given station and year: it is used to retain calibrations that allow for instrument characteristics and local site effects.

The first term on the right of equation (2) allows for the effect of solar zenith angle $\chi$ on the ionospheric conductivity over the station due to solar EUV and X-ray radiation and thus depends on the station's geographic latitude, the time-of-day and the time-of-year (see discussion at the end of this section about the choice of computing $\chi$ at the location of the magnetometer station). If the Sun is below the horizon, $\chi$ is set to ($\pi$/2): hence the coefficient $A$ controls the extent to which the effect of dayside conductivity at a given $\chi$ is enhanced over residual nightside values. Note that there are small changes to the precise formulation of Finch (2008), who used a $\cos^{0.5}(\chi)$ dependence, as predicted by Chapman production-layer theory and also used in a great many prior applications. However, Ieda et al. (2014) show that a conductivity dependence on $\cos^{0.7}(\chi)$ fits better with observations and is also predicted by theory when the upward gradient of the neutral atmospheric scale height is accounted for.

The second term on the right of equation (2) is the station's sensitivity due to its distance from the location of peak response, which is at an MLT of $T*$ in the midnight sector. The sine term in equation (3) is used to model the known earlier onset of enhanced substorm activity in summer (which is likely to also be a conductivity effect). Equation (3) yields $T*$ of 1 h MLT and 22 h MLT for the winter and summer solstices, respectively. This is based on the survey of mid-latitude station responses to substorm expansion phases by Finch (2008) and agrees well with the results of Liou et al. (2001), who found substorm onset was typically at $T = 22$ h in summer but 23.5 h in winter. Similar behaviour was deduced by Wang et al. (2007). We note that we are most interested in the MLT where auroral electrojet currents have peak effect on mid-latitude $K$ indices: this is close to, but not the same as, the MLT of onset (Clauer & McPherron, 1974; Chu et al., 2014).

Finch (2008) assumed that the factors $A$ and $B$ were constants and had considerable success in modelling the average response of different stations and indices. However, there are reasons to also think that the relative importance of the two terms in equation (2) might change systematically with the level of geomagnetic activity. Firstly, particle precipitation fluxes are higher during enhanced activity over a wide range of locations (including mid-latitudes; e.g., Shiokawa et al., 2005), which could lead to the relative contribution of photon-induced conductivity, and hence the dependence on $\cos^{0.7}(\chi)$, becoming weaker: hence the factor $A$ might be reduced at higher activity levels. Secondly, the auroral oval expands equatorward when activity is enhanced, making the second factor (associated with the spatial proximity of the auroral electrojet) more important. This could have a number of effects. The factor $B$ sets the amplitude of the diurnal variation seen by the station because of the variation in its proximity to the peak of the substorm current wedge. For these reasons the factors $A$ and $B$ are here treated as functions of the geomagnetic activity level.

Note that the Finch (2008) model employs the photon-induced conductivity above the station, which may not be the most appropriate location given that the majority of the current flows along the auroral oval in the auroral electrojet. We investigated this using three different locations at which the solar zenith angle $\chi$ was evaluated, namely: (A) the nominal auroral oval latitude at the same MLT as the station; (B) the location of the station; and (C) midway between these two. The goodness-of-fit metric (the root-mean-square (r.m.s.) deviation, $\Delta_{RMS}$) was very similar in all three cases (for the $aa$ index $\Delta_{RMS}$ was 0.122, 0.116, and 0.118 nT, for options (A), (B) and (C) respectively, whereas assuming the station sensitivity was a constant gave $\Delta_{RMS} = 0.132$) but the small differences give a preference ranking order of (B), (C), then (A). We here use option (B), largely because it avoids using a nominal latitude of the auroral oval rather than because it gives a better fit (the differences between the three cases being minimal and not statistically significant). This can be understood physically by thinking about the extremes of conductivity production in the auroral oval and considering the auroral electrojet to be linked to a pair of filamentary field-aligned currents (upward and downward at its westward and eastward ends, respectively) in the current wedge. If the conductivity were purely due to particle precipitation, the current along the oval would be a pure Cowling current (i.e., the Hall current is suppressed) along the oval where the precipitation is enhancing the conductivity. In this case, there would be no solar zenith angle dependence. If the conductivity were purely generated by solar photons, it would be enhanced both inside and outside the auroral oval. In this case, the Pedersen current (and therefore electric field and Hall current) would spread out in latitude from the line connecting the two filamentary currents (see, for example, Fig. 3 of Southwood, 1987). If the conductivity were spatially homogeneous, this spreading would be symmetric to the north and to the south of the oval; however, in reality it will be preferentially on the low-latitude side of the oval where $\chi$ is lower. Add to this the distance-squared decrease in the effect on the field at the station that is inherent in the Biot-Savart law, it is clear that the most relevant photon-induced conductivity would be equatorward of the oval and so closer to the station than the auroral oval. Hence using the location of the station (option B) is a reasonable way of quantifying the photon-enhanced conductivity effect.

# 3 Derivation of the coefficients $A$ and $B$ for sensitivity modelling of the $aa$ stations

To complete the set of equations used in this paper to compute $S_o$ for a given station at a given $F$, UT and year, we derive empirical expressions for $A$ and $B$, here quantifying the level of geomagnetic activity by the $aa$ index (after implementation of the corrections for the effect of the secular change in the geomagnetic field, as detailed in Paper 1) so that we can use the equations to correct all $aa$ values back to 1868.

Our approach is to assume that the spatial distribution of the $am$ observing stations is ideal, so we neglect any influence of limitations to the $am$ network on the UT-$F$ pattern shown in Figure 2e. This is an assumption, but as $am$ is by far the most homogeneous and most global range-based index that we have, it is an assumption that has been made, often tacitly, in a great number of previous studies. This being the case, the UT-$F$ pattern for the ratio of any index, divided by the simultaneous $am$ value, reveals (at least to first order) the UT-$F$ pattern for the sensitivity of that index, the solar forcing term in equation (1) having been cancelled.

Figures 3a–3c show the average UT-$F$ patterns for $aa'_N/am$, $aa'_S/am$ and $aa'/am$. The prime denotes that the correction for the secular drift, as developed in Paper 1, has been applied (so $aa'_N = \alpha_c \cdot aa_N/s(\delta) + \beta_c$, $aa'_S = \alpha_c \cdot aa_S/s(\delta) + \beta_c$ and $aa' = aa'_N + aa'_S)/2$, where $aa_N$ and $aa_S$ are the "classic" hemispheric $aa$ indices, $s(\delta)$ is the time-dependent scaling factor for the station in question, and $\alpha_c$ and $\beta_c$ are daisy-chained calibration factors that make all corrected $aa$ values consistent with the $am$ index for 2002–2009. Note also that in Paper 1, annual means of $aa'_N$, $aa'_S$ and $aa'$ were referred to as $aa_{HN}$, $aa_{HS}$ and $aa_H$, respectively. This nomenclature applies to the annual means because in the present paper we amend 3-hourly $aa'_N$, $aa'_S$ and $aa'$ values (to 3-hourly values we call $aa_{HN}$, $aa_{HS}$

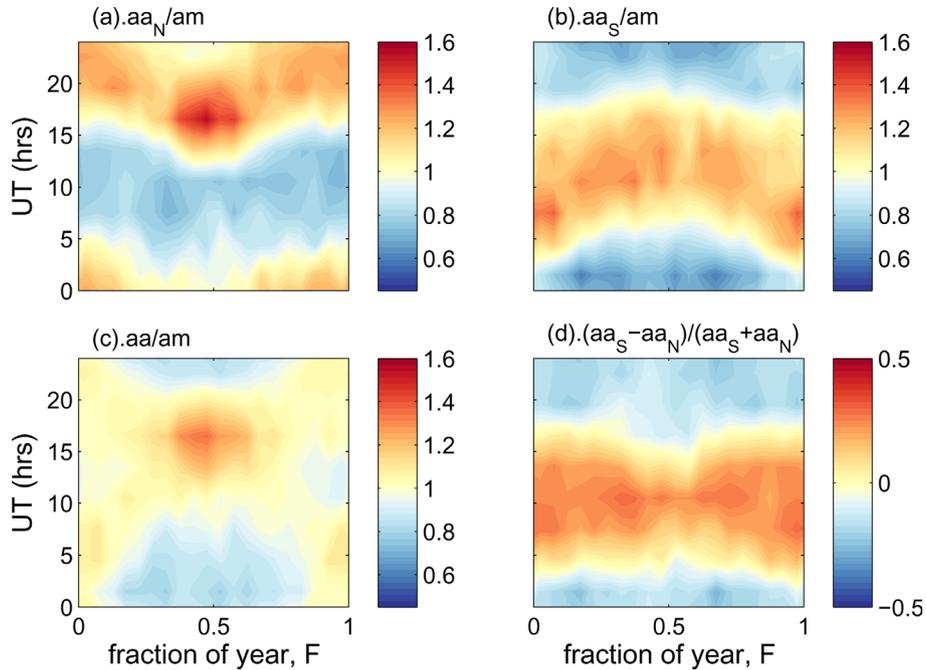

**Fig. 3.** Time-of-year ($F$), time-of-day (UT) plots of observed ratios of geomagnetic indices. All indices are 3-hourly so there are 8 UT bins and the 3-hourly ratios have been averaged into 20 equal-sized bins in $F$. (a) $aa'_N/am$; (b) $aa'_S/am$; (c) $aa'/am$; and (d) the $aa$ hemispheric anisotropy ratio $(aa'_S - aa'_N)(aa'_S + aa'_N)$. This plot is for all data in 1959–2017 with $70 \leq aa' < 110$ nT, for which the mean $aa'$ is 88.15 nT.

and $aa_H$) in such a way that their annual means remain unchanged. Figure 3d gives the pattern for the north-south anisotropy in $aa'$, $(aa'_N - aa'_S)/(aa'_N + aa'_S)$. We use all available data between 1959 and 2017 to keep the numbers of samples in each (UT, $F$) bin as high as possible. The data used to generate the example shown in Figure 3 are for all data points (since 1959) giving an $aa'$ index value that was relatively large (in the range $70 \leq aa' < 110$ nT, for which the mean $aa'$ is 88.15 nT).

Figure 4 shows the UT-$F$ patterns of the best-fit to Figure 3 of the modelled sensitivity from the implementation of the Finch (2008) model used here, as given by equations (2) and (3). In Figure 4, the solar zenith angle and MLT are evaluated for the relevant station location, UT and $F$ and for the year 1998 which is the midpoint of the data interval used in Figure 3. The coefficients $A = 0.11$ and $B = 0.28$ were derived by iteration using the Nelder-Mead search method to minimise the mean square of the deviation of all 640 pixels in Figure 4 from its corresponding pixel in Figure 3. The number 640 arises from the use of the 8 UT bins of the $aa$ and $am$ indices with our choice of 20 $F$ bins, so each panel contains 160 pixels and there are 4 panels. Note that all pixels in all four patterns are given equal weight by this procedure. The relatively low value of $A$ in this case means that the peak associated with the conductivity term in equation (2) is modest: this peak appears at the minima of the solar zenith angle $\chi$, which is near 17 UT and $F = 0.5$ for the northern hemisphere $aa$ station and near 7 UT and $F = 0$ (and hence also $F = 1$) for the southern hemisphere $aa$ station. The main effect in Figure 4 is the diurnal variation caused by the station's daily journey in MLT and hence the main feature is the UT variation in its sensitivity, the amplitude of that variation being set by $B$. Comparison of

Figures 3 and 4 show that model is capturing all the observed variations quite well.

At lower average $aa'$ values, the peak sensitivity at minimum solar zenith angle becomes much more pronounced. This can be seen in Figure 5, which is the same as Figure 3, but for the range $10 \leq aa' < 20$ nT (for which the mean $aa'$ is 14.42 nT). The best-fit model patterns for this case are shown in Figure 6, which are for $A = 0.58$ and $B = 0.33$.

It was possible to keep enough samples in each bin to see the average sensitivity patterns by dividing the full range of $aa'$ into 8 bins: $0 \leq aa' < 10$ nT; $10 \leq aa' < 20$ nT (the example presented in Figs. 5 and 6); $20 \leq aa' < 30$ nT; $30 \leq aa' < 40$ nT; $40 \leq aa' < 60$ nT; $50 \leq aa' < 90$ nT; $70 \leq aa' < 110$ nT (the example presented in Figs. 3 and 4); and $aa' \geq 100$ nT. Figure 7 gives a scatter plot of the modelled sensitivities in each of the 160 UT-$F$ pixels of the relevant pattern (using the best-fit $A$ and $B$ values for each $aa'$ bin) and all 8 $aa'$-bins (giving 1280 data points in total), as a function of the index ratio that they are fitted to. If the model were perfect, then all the points would lie on the blue line. It can be seen that the model has captured that trend. However, there is scatter around the line. We can compare the use of the modelled sensitivity factor to that assumed for the corrected $aa$, $aa'$, which is always unity ($S_o = 1$, which means that points would all lie on the diagonal red line). If we take the r.m.s. deviation of all the modelled sensitivities from observed ratios (1280 values, being 160 UT-$F$ pattern pixels in each of 8 $aa'$ range bins), $\Delta_{RMS}$, the ideal value would be zero in each case. (However, remember such a result would render all but one of the network of mid-latitude geomagnetic observatories redundant for space science studies as instead we could use the one station in conjunction with the model). For the $aa'_N$ index, assuming the sensitivity was constant at unity gives $\Delta_{RMS} = 0.232$, whereas using the fitted model value gives $\Delta_{RMS} = 0.143$. Therefore the model is reducing r.m.s. uncertainties (compared to not considering the station sensitivity) by 40% for $aa'_N$ (note that this analysis assumes there are no errors in the $am$ index).

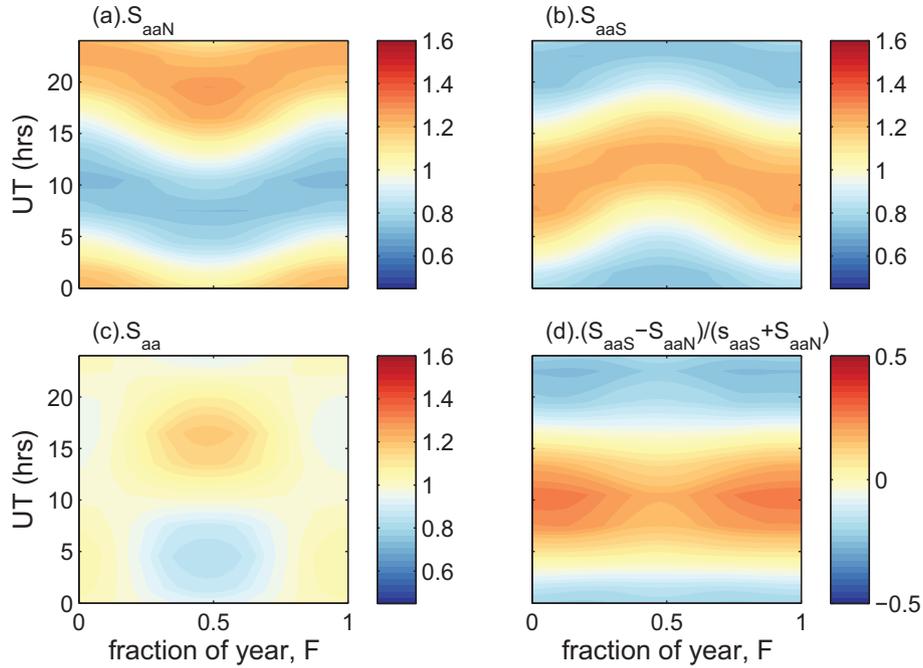

**Fig. 4.** Modelled UT-$F$ sensitivity patterns for $aa' = 88.15$ nT and so corresponding to the data shown in Figure 3. Sensitivities are modelled for the year 1988 at the same times as each 3-hourly $aa$ data point and the data then processed exactly as are the data in Figure 3 and plotted using exactly the same scales and colour scales: (a) $S_{aa_N}$; (b) $S_{aa_S}$; (c) $S_{aa}$; and (d) $(S_{aa_S} - S_{aa_N})(S_{aa_S} + S_{aa_N})$.

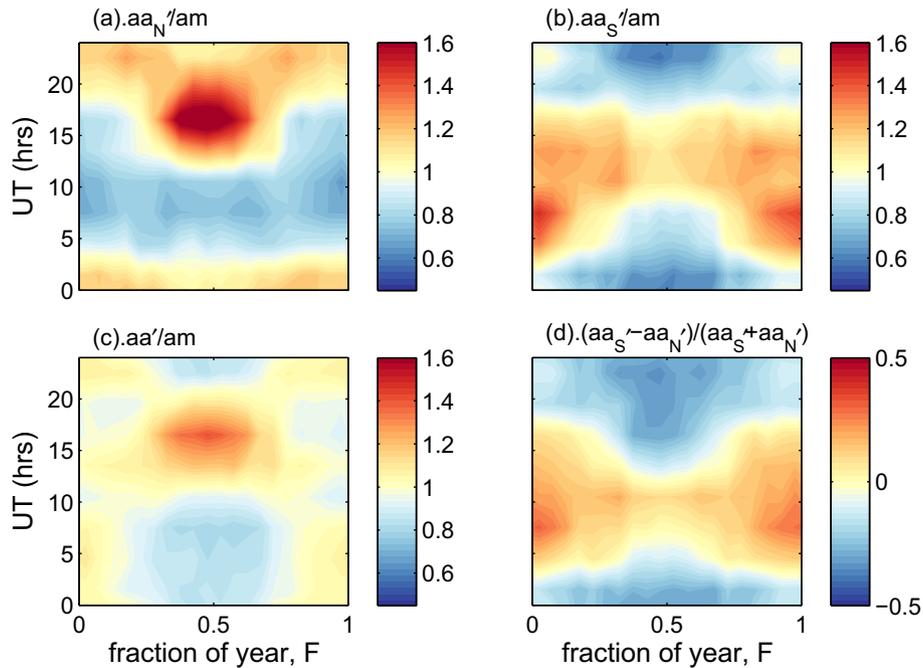

**Fig. 5.** Same as Figure 3, for $10 \leq aa' < 20$ nT, for which the mean $aa'$ is 14.42 nT.

This is the improvement in the mean for one bin of the UT-$F$ pattern, i.e., for one 3-hourly UT interval and averages in $F$ over 365/20 = 18.25 days. For the $aa'_S$ index, assuming the sensitivity was constant at unity gives a $\Delta_{RMS} = 0.265$, whereas using the fitted model value gives $\Delta_{RMS} = 0.174$. Therefore the model is reducing r.m.s. uncertainties (compared to not considering the station sensitivity) by 35% in this case. For the $aa'$ index, assuming the sensitivity was constant at unity gives a $\Delta_{RMS} = 0.132$, whereas using the fitted model value gives $\Delta_{RMS} = 0.116$. Therefore the model is reducing r.m.s. uncertainties (compared to not considering the station sensitivity) in $aa'$ by 11%. It is not surprising that the model improves the agreement with the $am$ pattern by much less for $aa'$, because the averaging of $aa'_N$ and $aa'_S$ to give $aa'$ is carried out precisely to also achieve this error reduction. These improvements are all quite modest. However, they are not the most important point. One of the key objectives in introducing the model is to bring the northern and southern hemisphere $aa$ indices into better agreement with each other, and so give us greater confidence that the average of the two is meaningful on timescales less than 1 year. If we consider the north-south anisotropy, $(aa'_N - aa'_S)/aa'_N + aa'_S)$, assuming $S_o = 1$ in both hemispheres gives a $\Delta_{RMS} = 0.989$, whereas using the fitted model value gives $\Delta_{RMS} = 0.112$. Therefore, in this case the model is reducing uncertainties (compared to not considering the station sensitivity) by 90%. The improvement is proportionally much greater in this case of the hemispheric anisotropy because there is self-consistent improvement to both $aa'_N$ and $aa'_S$. Hence the model gives improvements to both the hemispheric $aa'$ indices, but they are relatively modest (35–40%), and improvements are quite small for $aa'$ (~10%). However, the model can be very significant in reducing the asymmetry between northern and southern hemisphere indices.

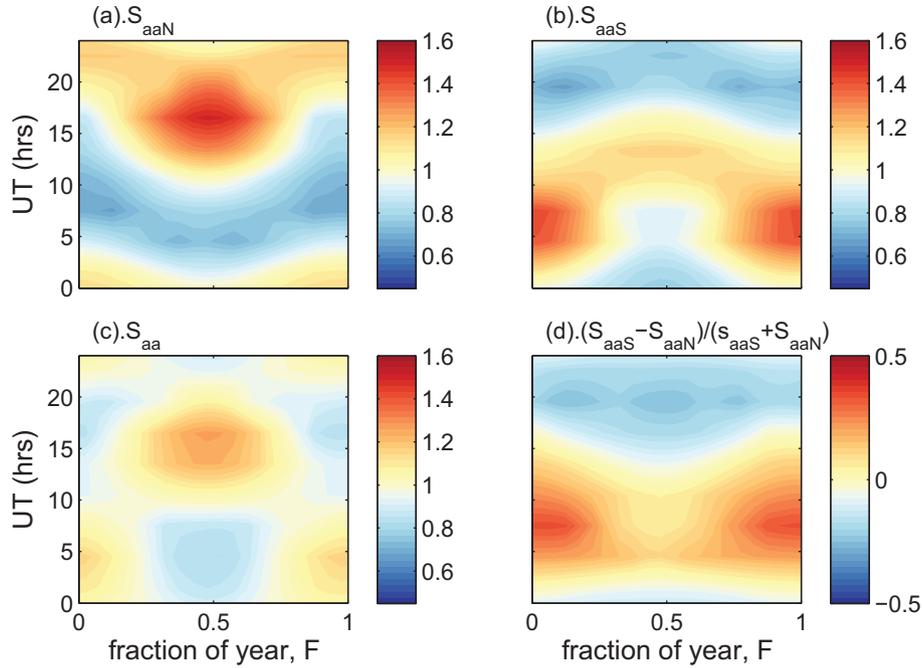

**Fig. 6.** Same as Figure 4, for $aa' = 14.42$ nT, and so corresponding to Figure 5.

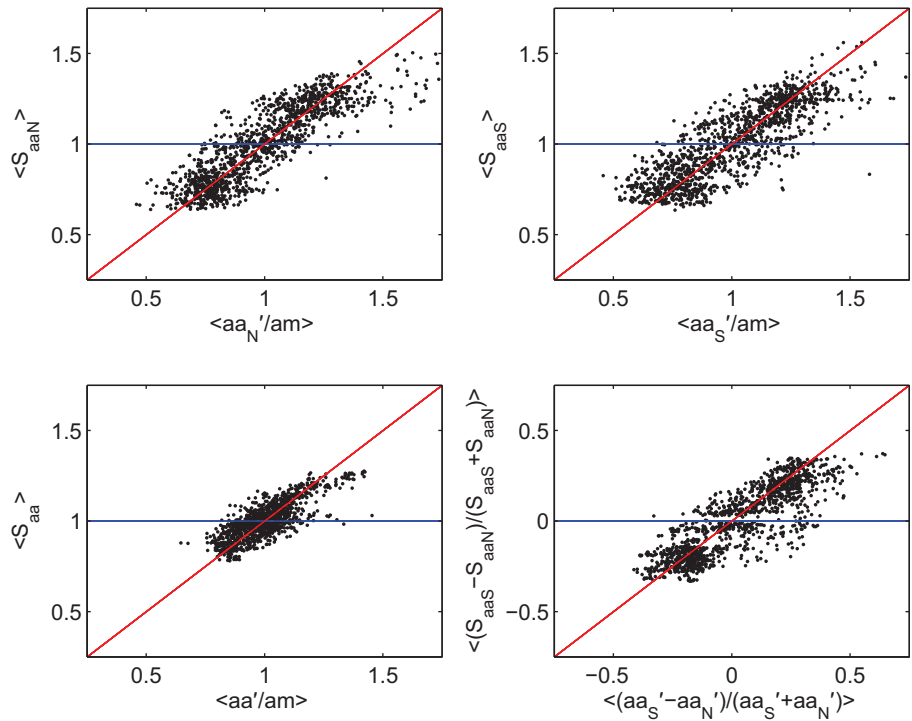

**Fig. 7.** Scatter plots of mean modelled sensitivity as a function of mean index ratio in each of the 160 UT-$F$ bins used in Figures 2–5. This plot is for all 8 bins in $aa'$ used to generate values of $A$ and $B$: $0 \leq aa' < 10$ nT; $10 \leq aa' < 20$ nT; $20 \leq aa' < 30$ nT; $30 \leq aa' < 40$ nT; $40 \leq aa' < 60$ nT; $50 \leq aa' < 90$ nT; $70 \leq aa' < 110$ nT; and $aa' \geq 100$ nT. Plots are for the modelled sensitivity against observed ratios for (a) $aa'_N/am$; (b) $aa'_S/am$; (c) $aa'/am$; and (d) the $aa'$ hemispheric anisotropy ratio $(aa'_S - aa'_N)(aa'_S + aa'_N)$. The red lines are where the model sensitivity map perfectly reproduces the index ratios and all points would lie on the horizontal blue line if the sensitivities were assumed constant.

The best-fit values of $A$ and $B$ for the 8 $aa'$ bins used are plotted as black dots and mauve circles, respectively, in Figure 8. These points are plotted at the mean $aa'$ value for the (overlapping) $aa'$ bins which are shown by the cyan and grey bars at the top of the plot. The figure shows how the conductivity term is relatively more important at low $aa'$ levels but the MLT term becomes more important at higher $aa'$ levels. The black and mauve lines are simple ad-hoc fits to the points, given by

$$A = 1.3e^{\{-aa'/15\}} + 0.1, \qquad (4)$$

$$B = 0.38(1 - \{(40 - aa')/70\}^2) \quad \text{for } aa' < 75.9 \, \text{nT},$$

$$B = 0.28 \quad \text{for } aa' \geq 75.9 \, \text{nT}, \qquad (5)$$

Equations (4) and (5) can be used to give the $A$ and $B$ coefficients for a given $aa'$ value, which can be used in equation (2) to compute the station sensitivity.

Note that the UT-$F$ patterns of $S_o$ will vary with secular change in the geomagnetic field because the MLT, $T$, of the station at a given UT will vary. As in Paper 1, we use a spline of the IGRF-12 (Thébault et al., 2015) and gufm1 (Jackson et al., 2000) models to predict the MLT at a given UT for each date and station, and so allow for this effect. This is an additional secular change to that considered in Paper 1 (associated with the latitude of the station and the closest proximity of the average auroral oval) but only influences the time-of-year/time-of-day variation around the annual mean and not the annual mean itself. The UT-$F$ maps of $S_o$ that we generate show that, over the centennial timescales of the $aa$ index, this can have a significant effect on the patterns of $S_o$ for a given station. The solar zenith angle calculation is made for a given site at the UT, $F$ and year at the centre of each of the 438,296 3-hourly $aa$ index intervals during 1868–2017, allowing for all variations in the Sun's declination.

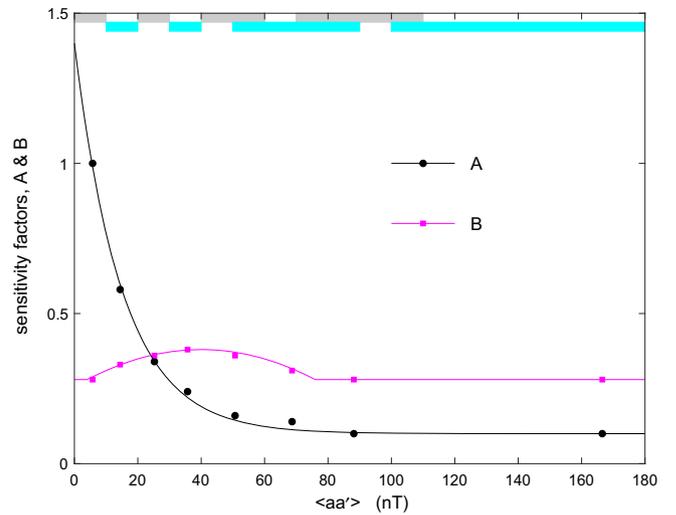

**Fig. 8.** The points show derived values of the best-fit coefficients $A$ (black dots) and $B$ (mauve squares) for the 8 $aa'$ bins as a function of the mean $aa'$ value for that bin. The black and mauve lines are the fits to these points given by equations (4) and (5), respectively.

In this paper, we use the computed sensitivities to correct 3-hourly $aa'_N$ and $aa'_S$ values (i.e., after time-dependent scaling factors $s(\delta)$ that allow for the effects of the secular drift in the geomagnetic field, as derived in Paper 1, have been applied). This gives corrected hemispheric indices:

$$aa_{HN} = f_N \cdot aa'_N/S_N, \qquad (6)$$

$$aa_{HS} = f_S \cdot aa'_S/S_S, \qquad (7)$$

where $S_N$ and $S_S$ are the station sensitivities (computed from Eqs. (2)–(5) for, respectively, the northern and southern hemisphere $aa$ station in use at the time. The factors $f_N$ and $f_S$ ensure that the annual (calendar-year) means derived in Paper 1 are not altered by the allowance for the variations of station sensitivity on timescales less than a year. Hence

$$f_N = \langle aa_{sN} \rangle_{\tau=1\,yr}/\langle aa'_N/S_N \rangle_{\tau=1\,yr}, \qquad (8)$$

And

$$f_S = \langle aa'_S \rangle_{\tau=1\,yr}/\langle aa'_S/S_S \rangle_{\tau=1\,yr}. \qquad (9)$$

Note that, even though $S_N$ and $S_S$ are normalised to be unity when averaged over all UT and times-of-year, the factors $f_N$ and $f_S$ will still, in general, differ from unity because of non-uniformity of activity occurrence within the year (for example, if in any one year, more geomagnetic activity happened, by chance, to occur when $S_N > 1$ than when $S_N < 1$, then $f_N$ will be less than unity). The homogenised hemispheric indices,

$aa_{HN}$ and $aa_{HS}$, are then averaged to give the corrected 3-hourly $aa$ index:

$$aa_H = (aa_{HN} + aa_{HS})/2. \tag{10}$$

## 4 Comparison of the new hemispheric indices

A measure of the degree of success of this procedure would be the extent to which the corrected $aa_{HN}$ and $aa_{HS}$ are similar, compared to the classic hemispheric indices, $aa_N$ and $aa_S$. Complete success would mean that $aa_{HN}$ and $aa_{HS}$ were identical (but, as noted above, this would also mean that the sensitivity model was so good that we could dispense with a full array of magnetometer stations and have just one, used in conjunction with the model). There is a limit to how much this approach can achieve. Consider a situation where $S_N$ is large (>1) and $S_S$ small (<1) such as around 20 UT and $F = 0.5$. Dividing $aa'_N$ by $S_N$ should give a reliable value of $aa_{HN}$, but if $S_S$ is small enough, the required signal may have fallen below the noise level and so dividing $aa'_S$ by $S_S$ (<1) increases both the noise and the signal and a reliable value of $aa_{HS}$ is not obtained. This effect was often noted at times of low activity when making comparisons of the $am$ index with the signal from a single station at a time when its $S_o$ value was low. The point is that an $am$ value is the average of the data from a number of stations which gives addition of the signal and cancellation of the noise and so $am$ has greater sensitivity to small fluctuations than does $aa$ or $aa'$. There are other limitations which are discussed in Section 6.

Figure 9 plots the occurrence of combinations of the classic $aa_N$ and $aa_S$ values in the upper panels and of the new corrected $aa_{HN}$ and $aa_{HS}$ values in the lower panels, with left-hand plots being for 3-hourly values and the right-hand plots being for daily means. The plot is for all available data (1868–2017). The diagonal mauve lines are the ideal case, with equal values for the two hemispheres. The pixels are logarithmically-sized and the number samples, $N$, in each pixel is color-coded. Figure 9a stresses the quantised nature of the classic $aa$ indices and that, especially at average and low $aa$, almost any value of $aa_S$ is possible for a given $aa_N$, and vice-versa. The correlation coefficient, $r$, between the full sequence (1868–2017) of 3-hourly $aa_N$ and $aa_S$ values is 0.66 and the r.m.s. deviation of the two from the $aa$ value, as a ratio of that $aa$ value, is $\delta = 0.53$. Figure 9c is the corresponding plot for the new indices, $aa_{HN}$ and $aa_{HS}$ and is very different in character. The values have been moved toward the diagonal and have become continuous in nature (although there is still some clustering of data points around the allowed combinations of the classic $aa$ indices). The correlation between $aa_{HN}$ and $aa_{HS}$ is increased to $r = 0.79$ and $\delta$ reduced to 0.45. For the daily means of the classic $aa$ indices, $Aa_N$ and $Aa_S$ (Fig. 9b), $r$ is 0.92 and $\delta$ is 0.28 and for the daily means of the new indices (Fig. 9d), $Aa_{HN}$ and $Aa_{HS}$, $r$ is (very slightly) increased to 0.93 and $\delta$ further reduced to 0.22. Hence the allowance for the station sensitivities has succeeding in increasing the agreement between the two hemispheric indices.

Figure 10 gives further comparisons of the hemispheric agreement for the classic and new $aa$ indices. The lower panel shows the coefficients of determination $r^2$ (where $r$ is the correlation coefficient) between southern and northern hemisphere indices, evaluated in calendar year intervals. The green and black lines are for the 3-hourly indices, green being for classic $aa$ indices ($aa_N$ and $aa_S$) and the black lines being for the new homogenized indices ($aa_{HN}$ and $aa_{HS}$). It can be seen that in all years the new indices have been brought into closer agreement with $r^2$ typically raised from around 0.4 to 0.55. The value is always higher for the new indices but there are a small number of years for which $r^2$ is high ($\approx 0.7$) in both the classic and the new indices: this appears to be a limit to how far the hemispheric indices can agree when they are compiled from just a single station. The red and blue lines are for daily mean data, red being for $Aa_N$ and $Aa_S$ and blue for $Aa_{HN}$ and $Aa_{HS}$. Again agreement is always slightly better for the new indices but the improvement is small for daily averages. The upper panel shows the annual means of $aa_H$ and it can be seen that the long term trend does not influence the hemispheric agreement. There is a tendency for the years of high agreement to occur one year before solar cycle minima. In all cases, there are slightly lower levels of agreement before 1880 which appears to indicate that there were increased measurement errors in one, or both, of the magnetometers at this time.

The left hand plots of Figure 11 compare the cumulative probability distributions (c.d.f.s) of the classic and new indices with the classic indices in the upper panel and the new indices underneath. The quantised nature of $aa_N$ and $aa_S$ (and to a lesser extent $aa$) can be seen in the upper panel. Note that larger southern hemisphere values are consistently more common than northern hemisphere values for $aa < 58$ nT, but the opposite is true for $aa > 58$ nT. These asymmetries in the classic $aa$ index distributions were pointed out by Bubenik & Fraser-Smith (1977) and by Love (2011). The lower panel shows the corresponding c.d.f.s for the new indices $aa_{HN}$, $aa_{HS}$ and $aa_H$. It can be seen that the new indices are essentially continuous, rather than quantised, and that the major asymmetry between the northern and southern distributions has been removed. The agreement of the $aa_{HN}$ and $aa_{HS}$ distributions is not perfect, but it is much better than for $aa_N$ and $aa_S$. The right-hand plots of Figure 11 show scatter plots of the old and new values. The upper panel shows $aa_{HN}$ as a function of $aa_N$ as red dots and $aa_{HS}$ as a function of $aa_S$ as blue dots. The vertical spreads reflect the range of sensitivity values applied. Careful inspection reveals that the corrections are not independent of the index value. For example at large $aa_N$, $aa_{HN}$ is more often reduced compared to $aa_N$ rather than the other way round. In other words, high $aa_N$ tends to be recorded at times of high station sensitivity, $S_N$, which amplifies the value detected. However, this is not always the case and the plot also shows cases where high $aa_N$ was recorded at times of $S_N < 1$. The tendency is reflected in the change to the CDF. The lower-right plot is the scatter plot between $aa_H$ and $aa$. The increased effect of the conductivity term in the sensitivity model at low $aa$ can be seen as a very slight non-linearity in the plot. In general, $aa_H$ values are lower than $aa$ by between about 0 and 30%. The average decrease of about 15% is mainly due to the calibration of the new indices against $am$ data over the interval 2002–2009 (see Paper 1).

Quantile-quantile (q-q) plots are a standard method for testing if two populations share the same form of distribution, because points lie along the diagonal if they do. Figure 12a

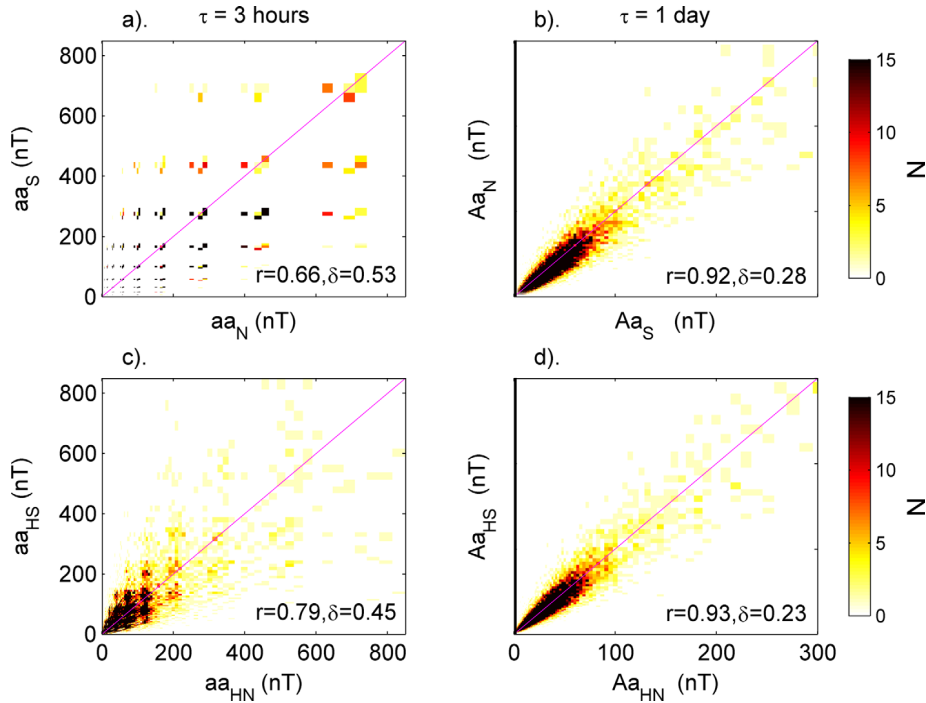

**Fig. 9.** The difference between the northern and southern hemisphere *aa* indices. The upper two panels are for the classic *aa* indices, the bottom two panels after correction for secular geomagnetic drift and division by the station sensitivity factors. The left-hand panels are for 3-hourly values, the right-hand panels are for daily means. In each panel the number of samples *N* is colour-coded as a function of the southern index value (vertical axis) and the northern index value (horizontal axis) and the mauve line is the ideal case where the two are the same for all samples. Note that pixel sizes are increased logarithmically with increased index values. (a) 3-hourly values of classic *aa* indices, $aa_N$ and $aa_S$. (b) Daily means of classic *aa* indices, $Aa_N = \langle aa_N \rangle_{\tau=1\,day}$ and $Aa_S = \langle aa_S \rangle_{\tau=1\,day}$ (c) Corrected 3-hourly indices, $aa_{HN}$ and $aa_{HS}$ derived using equations (6)–(9). (d) Daily means of corrected indices $Aa_{HN} = \langle aa_{HN} \rangle_{\tau=1\,day}$ and $Aa_{HS} = \langle aa_{HN} \rangle_{\tau=1\,day}$. Each panel gives the correlation coefficient, *r*, between the full sequence (1868–2017) of northern and southern hemisphere values and the r.m.s. fractional deviation of the two from the *aa* value, δ. For $aa_N$ and $aa_S$, *r* is 0.66 and δ is 0.53; for $aa_{HN}$ and $aa_{HS}$, *r* is increased to 0.79 and δ reduced to 0.45. For $Aa_N$ and $Aa_S$, *r* is 0.92 and δ is 0.28; for $Aa_{HN}$ and $Aa_{HS}$, *r* is (very slightly) increased to 0.93 and δ reduced to 0.22. Note the quantised nature of the classic $aa_N$ and $aa_S$ values in (a), but also the splitting of the allowed values caused by the use of scaling factors to intercalibrate the stations.

is the q-q plot for 3-hourly values of $aa_N$ and $aa_S$: the quantization of $aa_N$ is evident, and the scatter of points away from the green diagonal show distributions are not closely matched at all levels, as also shown by Figure 11a. Figure 12c is for 3-hourly values of $aa_{HN}$ and $aa_{HS}$. It can be seen these distributions are continuous and similar up to the 99.97 percentile (the orange point). For the largest 0.03% of 3-hourly values (above the orange point) we see some divergence of the two distributions with the occurrence of large events being slightly lower for the southern hemisphere (although the distributions agree around the 99.99 percentile). Figures 12b and 12d are the same comparison for daily mean values. Above about the 99.5 percentile, the tails of the $Aa_N$ and $Aa_S$ distributions are not generally well matched, with quantiles for $Aa_S$ slightly, but persistently, at lower values than for $Aa_N$ although they do agree better near the 99.97 percentile (the orange point). For daily means $Aa_{HN}$ and $Aa_{HS}$, the distributions agree well all the way up to the 99.97 percentile but they disagree above this percentile with quantiles for $Aa_{HS}$ again persistently at lower values than for $Aa_{HN}$. Thus the homogenization has resulted in the northern and southern index distributions being of more similar shape, except for the extreme values above the

99.97 percentile (the orange points), which is where the rarity of events is likely to make their occurrence in the two hemispheric indices more dissimilar. As discussed in Section 5, there are combinations of UT and MLT when the southern station records lower values, possibly because of a UT variation in geomagnetic activity, but at no time is the northern station subject to this effect (because it is at a different longitude). Hence the divergence of the extreme event tails of these q-q plots (with generally fewer events seen in the southern hemisphere) appears to be a real physical effect, associated with the longitudes of the stations, and not due to measurement error and noise.

## 5 The time-of-day/time-of-year pattern for the new indices

Figure 13 compares the time-of-day/time-of-year (UT-*F*) patterns of the classic and new *aa* indices against that for the *am* index. Figure 13a shows the pattern for *am*, and reveals the quasi-equinoctial pattern discussed in Section 1. Figure 13b

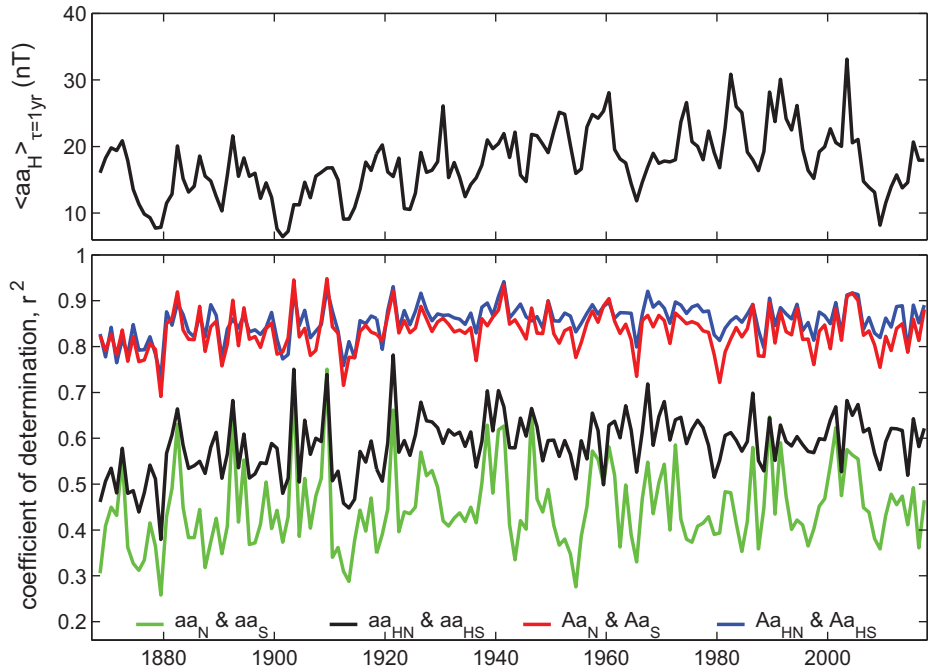

**Fig. 10.** (a) Annual means of the corrected *aa* index, $aa_H$. (b) The coefficients of determination ($r^2$, where $r$ is the correlation coefficient) for 1-year sequences of northern and southern hemisphere indices: (green) the classic *aa* indices $aa_S$ and $aa_N$; (black) the corrected aa indices $aa_{HN}$ and $aa_{HS}$; (red) daily means of the classic *aa* indices, $Aa_S$ and $Aa_N$; and (blue) daily means of the corrected *aa* indices $Aa_{HN}$ and $Aa_{HS}$.

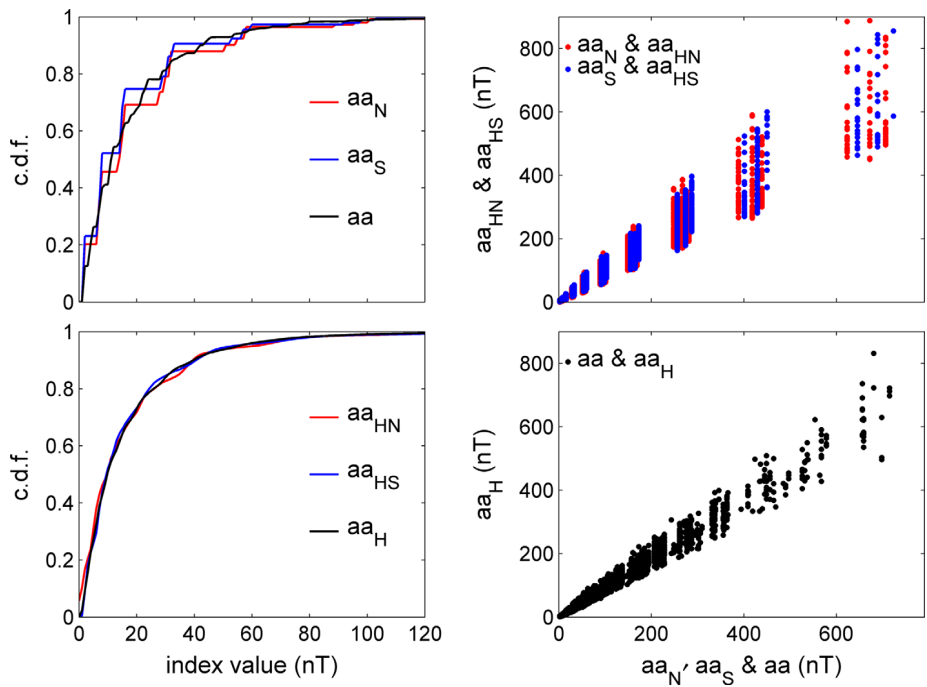

**Fig. 11.** (Left) Cumulative probability distributions (c.d.f.s) of indices and (right) scatter plots of uncorrected and corrected 3-hourly values. (a) c.d.f.s of classic *aa* indices $aa_N$ (in red), $aa_S$ (in blue), and *aa* (in black): the steps for the hemispheric indices show their quasi-quantised nature and note that larger southern hemisphere values are consistently more common than northern hemisphere values for *aa* < 58 nT, but the opposite is true for *aa* > 58nT. (b) Scatter plots of $aa_{HN}$ against $aa_N$ (red points) and of $aa_{HS}$ against $aa_S$ (blue points). (c) c.d.f.s of corrected *aa* indices $aa_{HN}$ (in red), $aa_{HS}$ (in blue), and $aa_H$ (in black). (d) Scatter plot of $aa_H$ against *aa*.

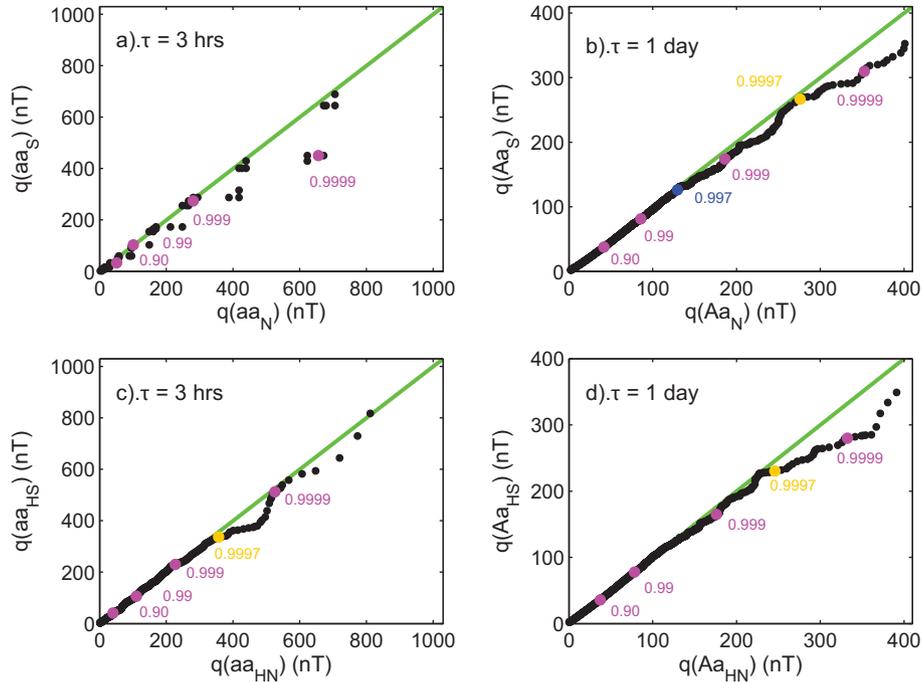

**Fig. 12.** Quantile-quantile ("q-q") plots to test the similarity of corresponding distributions for the southern and northern hemispheres: (a). for 3-hourly values of $aa_N$ and $aa_S$; (b) for the daily means of $aa_N$ and $aa_S$, $Aa_N$ and $Aa_S$; (c) 3-hourly values of $aa_{HN}$ and $aa_{HS}$ and (d) their daily means $Aa_{HN}$ and $Aa_{HS}$. The mauve points mark the 90%, 99%, 99.9% and 99.99% percentile points and the orange is for 99.97%. Also shown in blue in part (b) is the 99.7% percentile. Points would lie along the diagonal green line if the distributions from the two hemispheres had the same shape.

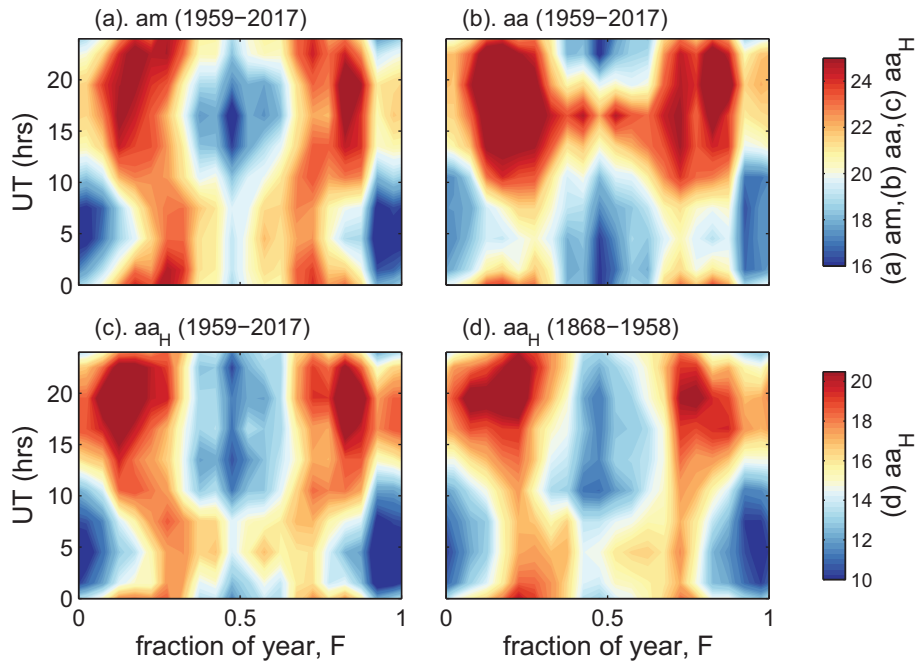

**Fig. 13.** Time-of-year ($F$)/time-of-day (UT) plots of mean geomagnetic indices. The means are for 8 bins UT and 20 equal-length bins in $F$. (a) is for the $am$ index, which covers the interval 1959–2017 (inclusive); (b) is for the classic $aa$ index, $aa$, for the same interval (1959–2017); (c) is for the corrected $aa$ index, $aa_H$, for the same interval (1959–2017); and (d) is for $aa_H$ for the prior interval 1868–1958. Parts (a)–(c) use the same (upper) colour scale, but because average values are lower before 1859, a slightly different scale is used in (d), given by the lower colourbar. The $am$ index and the corrected $aa$ index, $aa_H$, both show the "equinoctial" pattern, but the classic $aa$ index does not.

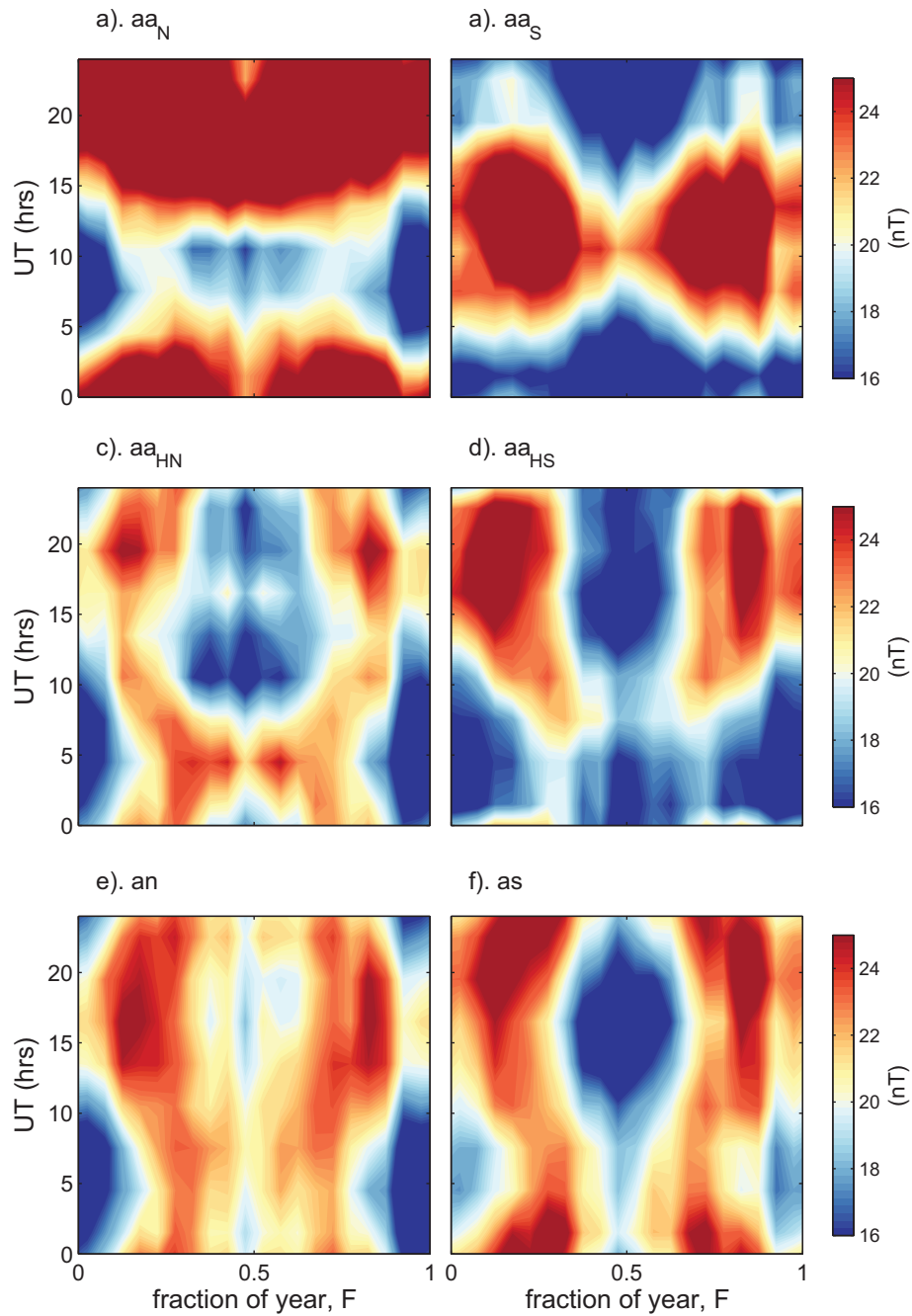

**Fig. 14.** Time-of-year (*F*)/time-of-day (UT) plots of hemispheric geomagnetic indices for 1959–2017. The means are for 8 bins UT and 20 equal-length bins in *F*. (a) Shows the variation for the classic northern hemisphere *aa* index, *aa*$_N$ and (b) is for the classic southern hemisphere *aa* index, *aa*$_S$. (c) and (d) are for the homogenised northern and southern (respectively) hemisphere *aa* indices, *aa*$_{HN}$ and *aa*$_{HS}$, which include the correction of the secular variation described in Paper 1 (Lockwood et al., 2018) and allowance for the variation of the station sensitivity developed in the present paper. (e) and (f) are for the northern and southern (respectively) hemisphere *am* indices, *an* and *as*.

shows the pattern for the classic *aa* index for the same years as are available for *am* (1959–2017): it can be seen that the spurious diurnal variation caused by having just one station in each hemisphere has seriously disrupted the pattern, with a marked minimum in the response in *aa* at 1–8 UT, and an excessive response at 12–23 UT that appears more axial than equinoctial in form. Figure 13c shows the pattern for *aa*$_H$ for

1959–2017. It can be seen that this pattern in the new index is more equinoctial and quite similar to that for *am*. This means that the shrinking of the difference between the annual variations of the new hemispheric indices, seen in Figure 1e and the flattening of the UT variation in Figure 1f have been achieved in a self-consistent way in the new homogenized *aa* indices. Figure 13d shows the UT-*F* pattern for *aa*$_H$ for all years before the start of the *am* index, i.e., for 1868–1958. It can be seen that it too shows an equinoctial-like pattern. As discussed in Section 1, we are not yet certain of the physical origin of the equinoctial pattern but none of the proposed mechanisms offer any reason why it should not be present before 1959 as well as after and Figure 12d shows that *aa*H reveals that it is. Note that the colour scale on which *aa*H is plotted in Figure 13d has been reduced by the ratio of average *aa*H values before and after 1959.

Figure 14 shows a much more stringent and revealing test of the new indices by looking to see if the equinoctial pattern is present in the new hemispheric indices on their own. Figures 14a and 14b show the UT-*F* patterns for the classic *aa* indices (respectively, *aa*N and *aa*S). It can be seen that the pattern is dominated by the MLT variation of the station in both cases, with strong peaks at all times of year around 21 UT in the *aa*N and around 11 UT in *aa*S. The effect of the semi-annual variation in the solar wind forcing of geomagnetic activity can also be seen, with peaks at the equinoxes, but the pattern is very far from equinoctial. Figures 14c and 14d show the UT-*F* patterns for the new homogenised hemispheric *aa* indices (respectively, *aa*HN and *aa*HS). It can be seen that, remarkably, the equinoctial pattern has partially emerged in both cases, although in neither case is the variation for *am* (shown in Fig. 12a) perfectly reproduced. Figures 14e and 14f show the UT-*F* patterns for the hemispheric *am* indices (respectively, *an* and *as*). The equinoctial pattern is again seen but, as for *aa*HN and *aa*HS, neither is an exact replica of the *am* variation. Some of the anomalous features in the new patterns for the new *aa* indices are seen in the hemispheric *am* indices: for example, the minimum in the response of the southern hemisphere indices at around 5 UT is present in both *as* and *aa*HS. Figure 14d is interestingly consistent with the results of Chambodut et al. (2013) who found the equinoctial pattern in *am* almost disappeared in the noon MLT sector at roughly 0–9 UT: the southern hemisphere *aa* station is at about 10.6–19.6 MLT in this interval and so has passed through the noon sector. Thus the loss of the equinoctial pattern in the *aa*HS data occurs at the time that we would expect from the results of Chambodut et al. (2013). On the other hand, the northern hemisphere *aa* station is at 23.7–8.7 h MLT in the 0–9 UT interval and so in the midnight and dawn sectors, for which Chambodut et al. (2013) strongly detect the equinoctial pattern at all UT. Hence we see no such gap in the equinoctial pattern in the *aa*HN. This appears to reflect a genuine UT variation in solar wind-magnetosphere coupling and/or in the response of the magnetosphere-ionosphere-thermosphere system. This is an issue that we will return to in a later paper.

# 6 Conclusions

By using a model of the sensitivity of a geomagnetic observing site that takes account of its solar zenith angle and its MLT (and to a small extent the geomagnetic activity level), we have generated a new "homogenised" data series of 3-hourly and daily-mean values: *aa*HN, *aa*HS and *aa*H. These also make use of the long-term recalibration of stations and the allowance for the change in the geomagnetic field that was implemented in Paper 1. The new indices are generated by a fixed algorithm that is the same for all magnetometer stations and show a number of improvements over the classic *aa* indices, namely:

1. The long-term drift of the northern and southern hemisphere indices is the same (see Paper 1).
2. The distributions of index values are continuous and not quantised.
3. The distributions of values for the northern and southern hemisphere indices are very similar.
4. The differences between simultaneous 3-hourly northern and southern hemisphere index values is reduced.
5. The correlation between the 3-hourly northern and southern hemisphere index values is increased (overall and in all individual years).
6. The correlation between the daily means of northern and southern hemisphere index values is slightly increased (overall and in all individual years).
7. The mean annual variation in the northern and southern hemisphere indices is very similar and that difference is consistent with a seasonal effect only.
8. The difference between the mean diurnal variations in the northern and southern hemisphere indices is greatly reduced and there is almost no residual diurnal variation in the new *aa*H index.
9. The equinoctial time-of-day/time-of-year pattern in the new *aa*H index matches that in *am*.
10. The equinoctial time-of-day/time-of-year pattern appears in the new hemispheric indices (but does not exactly replicate those in the hemispheric *am* indices).

We note that there are limits to how much a station sensitivity model can do in terms of correcting a one-station hemispheric range index such as *aa*N and *aa*S into a more representative global index made from a longitudinal ring of stations, such as *an* and *as*. As discussed above, one reason is that sensitivity at one station could be low enough for the signal to fall below the noise level. In such cases, dividing by the low (<1) sensitivity amplifies both the noise and the signal and will not recover the signal that was seen in the other hemisphere. Furthermore, Caan et al. (1978) found that, in addition to the amplitude of the response to a substorm varying with the MLT of the station, the waveform of the response varies also. This means that some substorms will cause a large range measurement in one 3-hour interval, but another station, at a different MLT, might detect the largest range measurement in the previous or the next 3-hour interval. Correction using division by the modelled sensitivity could not correct for such an occurrence. Hence the method has its limitations. However, in all the ways that we have tested the new homogenized indices, they out-perform the classic *aa* indices and so application of the sensitivity model has made improvements. Essentially, factors that ideally would average out when taking the mean of the two hemispheric indices but in practice do not exactly cancel, have here been allowed for, at least to some extent. The fact that there is a UT range when, because of its longitude, the southern hemisphere station sees lower values and this never occurs for the northern hemisphere station, will have been accounted for in average values of our new index because we calibrated both hemispheric data series against the *am* index.

There are two possible objections to using the modelled sensitivity to improve *aa* to *aa*$_H$ that we can foresee. The first is that we are correcting scaled and quantized *K* values that were generated against a fixed (*K*) scale. However, we note that this is already done in the generation of the classic *aa* indices because of the use of the station scaling factors. For the classic *aa*, these are constants for the station location, whereas here we are using scale factors based on the station location but that change with time because of secular change in the geomagnetic field and because of Earth's orbit and rotation. We allow for such effects using a repeatable algorithm that can be evaluated by anyone. The second objection is that we are using model values to adjust observations. Again, the principle of this is already inherent in the classic *aa* index because the range thresholds that define the *K*-index bands are set by a model (specifically, they are set by a model of the dependence of the observed range value on the separation of the station and the auroral oval). In addition, we note that the *IHV* geomagnetic index (Svalgaard & Cliver, 2007) assumes the equinoctial model of the UT-*F* dependence in its construction. Hence we think that in neither case are we doing something that is not already inherent in the generation of the classic *aa* index – we are just implementing a more complex scheme to correct for the limitations of the original *K*-index scaling.

In a later paper we will study how these station sensitivity considerations influence more complex range indices (such as *ap*, *Kp*, *am*, *an* and *as*) and how they have been influenced by changes to the network of stations from which they have been compiled. In the present paper we have restricted our attention to the *aa* index and through the ten improvements listed above, shown that applying the sensitivity model to the stations can greatly improve how the index quantifies geomagnetic activity, even though it only employs two stations. Note that in correcting the *aa* data series we have allowed for the effects of observatory location on sensitivity (including the observatory changes), secular drifts in the geomagnetic field and the intercalibration of the instrumentation and local site characteristics at the different observatories.

In another subsequent paper we will study large and extreme events in both the daily and 3-hourly data of the homogenised index. The 3-hourly classic *aa* values have been used to rank historic geomagnetic storms since 1868 by Vennerstrøm et al. (2016), but the rank order for the new 3-hourly *aa*$_H$ values vary considerably from that for the 3-hourly classic *aa* values.

The annual means of the new indices (*aa*$_{HN}$, *aa*$_{HS}$ and *aa*$_H$) are supplied in the supplementary material attached to Paper 1. The three-hourly values of the new indices (*aa*$_{HN}$, *aa*$_{HS}$ and *aa*$_H$) and their daily averages (*Aa*$_{HN}$, *Aa*$_{HS}$ and *Aa*$_H$) are supplied in separate files in the Supplementary material attached to the present paper. For some applications it may be useful to use 3-hourly or daily indices that have been corrected for the effects of secular field change and recalibrated but have not been further modified using our sensitivity model: these are also given in the supplementary material (3-hourly *aa*/*s*, *aa*$_N$/*s* and *aa*$_S$/*s* values and their daily means $\langle aa/s \rangle$, $\langle aa_N/s \rangle$, and $\langle aa_S/s \rangle$).

## Supplementary material

Supplementary material is available at https://swsc-journal.org/10.1051/swsc/2018044/olm

*Acknowledgements.* The authors are grateful to the staff of The International Service of Geomagnetic Indices (ISGI), France and collaborating institutes for the compilation and databasing of the *am* index which were downloaded from http://isgi.unistra.fr/data_download.php. We also thank the staff of Geoscience Australia, Canberra for the southern hemisphere *aa*-station *K*-index data, and colleagues at British Geological Survey (BGS), Edinburgh for the northern hemisphere *aa*-station *K*-index data. The work at University of Reading (UoR) is supported by the SWIGS NERC Directed Highlight Topic Grant number NE/P016928/1 with some additional support from STFC consolidated grant number ST/M000885/1. The work at École et Observatoire des Sciences de la Terre (EOST) is supported by CNES, France. Initial work for this paper was carried out by IDF as part of his PhD studies at Southampton University, where he was a part-time student: we are grateful to Rutherford Appleton Laboratory and to Southampton University for supporting that PhD work. The editor thanks two anonymous referees for their assistance in evaluating this paper.

---